\documentclass[journal]{IEEEtran}

\usepackage{graphicx} 
\usepackage{epstopdf}
\usepackage{textcomp} 
\usepackage{array}
\usepackage{cite}
\usepackage{amsmath}
\usepackage{booktabs} 
	\newcolumntype{T}{>{\footnotesize}c}
\ifCLASSINFOpdf
\else
\fi
\begin{document}
%
\title{Numerical Modeling of 3.5~\textmu m Dual-Wavelength Pumped Erbium Doped Mid-Infrared Fiber Lasers}
%
%
%


\author{
	Andrew Malouf,
	Ori Henderson-Sapir, \IEEEmembership{Member,~IEEE},
	Martin Gorjan,
	David J. Ottaway%
	\thanks{Manuscript received June xx, 2016; revised June xx, 2016. This work was supported by the South Australian Premier Research and Industry Fund (PRIF) Grant.}
	\thanks{A. Malouf, O. Henderson-Sapir, and D. Ottaway are with The University of Adelaide, Adelaide, S.A. 5005, Australia (email: andrew.malouf@adelaide.edu.au; ori.henderson-sapir@adelaide.edu.au; david.ottaway@adelaide.edu.au).}
	\thanks{M. Gorjan is with Spectra-Physics, 6830 Rankweil, Austria (e-mail: martin.gorjan@spectra-physics.at).}}

%
%

\markboth{IEEE Journal of Quantum Electronics,~Vol.~xx, No.~x, June~2016}{Malouf \MakeLowercase{\textit{et al.}}: Modeling and Optimization of Erbium Doped Mid-Infrared Fiber Lasers}
%



\maketitle

\begin{abstract}
The performance of mid-infrared Er\textsuperscript{3+}-doped fiber lasers has  dramatically improved in the last few years. In this paper we present a numerical model that provides valuable insight into the dynamics of a dual-wavelength pumped fiber laser that can operate on the 3.5~\textmu m and 2.8~\textmu m bands. This model is a much needed tool for optimizing and understanding the performance of these laser systems. Comparisons between simulation and experimental results for three different systems are presented.
\end{abstract}


%
\IEEEpeerreviewmaketitle


\section{Introduction}
%
%
%
%

\IEEEPARstart{N}{ew} mid-infrared laser sources will enable significant advances in a wide range of applications including spectroscopy \cite{schliesser2012mid}, remote sensing \cite{walsh2016mid}, non-invasive medical diagnosis \cite{kim2010potential}, and defense countermeasure \cite{molocher2005countermeasure}. The mid-infrared spectral range is of particular interest in molecular spectroscopy \cite{vainio2016mid}, since strong absorption features of many molecules (including hydrocarbons) are found there. The fundamental absorption lines for molecules containing bonds between hydrogen and carbon, nitrogen or oxygen are located between 2.8~\textmu m and 4~\textmu m \cite{vainio2016mid}. 

Dual-wavelength pumping (DWP) of an Er\textsuperscript{3+}-doped ZBLAN fluoride glass fiber is an efficient method of enabling a 3.5~\textmu m laser at room temperature. This method takes advantage of long-lived excited states that cause bottlenecks which normally limit laser performance \cite{hendersonsapir2014midinfrared}. Recent work has demonstrated that emission between 3.3~\textmu m and 3.8~\textmu m can be achieved on the $\mathrm{^{4}F_{9/2}\rightarrow^{4}I_{9/2}}$ transition when DWP is used \cite{hendersonsapir2016versatile}.

A numerical model has been developed to provide valuable insight into laser performance, the significance of competing processes, and the interactions that occur at the atomic and photonic level. The model can be used to analyze system design and predict optimum fiber specifications.

Several numerical models have been developed to study lasers that operate on the 2.8~\textmu m transition in Er\textsuperscript{3+}-doped fiber lasers. Gorjan \textit{et al.} \cite{gorjan2011role} numerically investigated the significance of interionic processes in Er\textsuperscript{3+}-doped ZBLAN while Li \textit{et al.} \cite{li2014modeling} numerically optimized parameters such as doping concentration. We present a model that has been developed and experimentally validated against three published DWP systems \cite{hendersonsapir2014midinfrared,hendersonsapir2016versatile,fortin2016watt} operating on the 3.5~\textmu m transition. We discuss the factors that limit their performance and methods of optimization. The model can be adapted to any fiber laser system.

The paper is organized as follows: Section II introduces the scientific and mathematical basis of the numerical model. Section III describes the validation procedure for the model. In Section IV we discuss the findings and potential optimizations. Finally, conclusions are presented in Section V.


\section{Basis of Numerical Model}

\begin{figure}[!h]
	\centering
	\includegraphics[width=1\columnwidth]{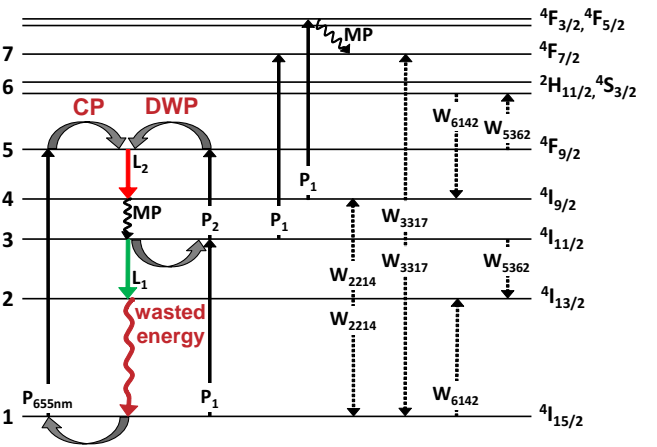}
	\caption{The energy levels of erbium ions showing DWP transitions due to pump absorption of the first ($\mathrm{P_{1}}$) and second ($\mathrm{P_{2}}$) pumps, 2.8~\textmu m lasing ($\mathrm{L_{1}}$), 3.5~\textmu m lasing ($\mathrm{L_{2}}$), and energy transfer processes $\mathrm{W_{ijkl}}$. A comparison between conventional pumping (CP) and dual-wavelength pumping (DWP) techniques is illustrated on the left. Two important multi-phonon (MP) decay processes are illustrated while other decay processes are omitted for brevity.}
	\label{fig:Erbium-energy-levels-with-DWP}
\end{figure}

The energy level transitions in Er\textsuperscript{3+} that are associated with the DWP system are illustrated in Fig. \ref{fig:Erbium-energy-levels-with-DWP} \cite{hendersonsapir2016versatile,hendersonsapir2016new}. Most decay processes are omitted for brevity but would simply be represented by arrows connecting each excited state to each level below it. A comparison of conventional pumping (CP) \cite{tobben1992room} and DWP \cite{hendersonsapir2014midinfrared} techniques used to generate 3.5~\textmu m lasing ($\mathrm{L_2}$) are included in the left of Fig. \ref{fig:Erbium-energy-levels-with-DWP}. The CP technique pumps ions from ground level $\mathrm{^{4}I_{15/2}}$ (1) directly to the upper laser level $\mathrm{^{4}F_{9/2}}$ (5) using one 655~nm pump source ($\mathrm{P_{655nm}}$).

The DWP technique uses a pump source with a wavelength of 966-985~nm \cite{hendersonsapir2016versatile,fortin2016watt} ($\mathrm{P_{1}}$) to excite ions from the ground level $\mathrm{^{4}I_{15/2}}$ (1) to level $\mathrm{^{4}I_{11/2}}$ (3), and a pump source with a wavelength of 1973-1976~nm \cite{hendersonsapir2016versatile,fortin2016watt} ($\mathrm{P_{2}}$) to excite ions from levels $\mathrm{^{4}I_{11/2}}$ (3) to $\mathrm{^{4}F_{9/2}}$ (5). DWP takes advantage of the long lifetime of level $\mathrm{^{4}I_{11/2}}$ (3) which acts as an elevated ``virtual" ground state and cycles ions between levels $\mathrm{^{4}I_{11/2}}$ (3) and $\mathrm{^{4}F_{9/2}}$ (5). DWP significantly increases efficiency compared with CP since less energy is wasted by decay from level $\mathrm{^{4}I_{11/2}}$ (3) back to the ground state. Low $\mathrm{P_2}$ powers enable lasing at 2.8~\textmu m ($\mathrm{L_1}$) on the $\mathrm{^{4}I_{11/2}\rightarrow^{4}I_{13/2}}$ transition.

A numerical model, titled ``Fiber Laser Atomic and Photonic Populations" (FLAPP), was developed in MATLAB \cite{MATLABR2013a} to solve the rate equations listed in Section \ref{sec:Rate-equations}. This model is mathematically similar to that described by Gorjan \textit{et al.} \cite{gorjan2011role}. 

\begin{figure}[!h]
	\centering{}\includegraphics[width=1\columnwidth]{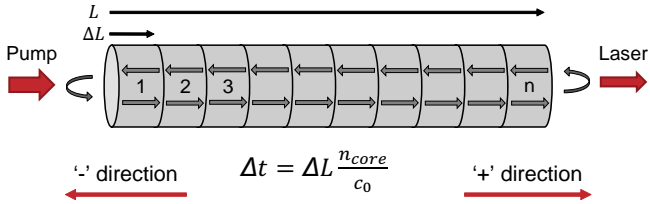}
	\caption{Numerical iteration of photon propagation in a fiber divided into $n$ length elements. Photons propagate in the `$\mathrm{+}$' or `$\mathrm{-}$' direction. $n_{core}$ is the refractive index of the fiber core and $c_{0}$ is the speed of light in a vacuum. The time step $\varDelta t$ is defined as the time required for light to traverse a single fiber element of length $\varDelta L=\frac{L}{n}$.}
	\label{fig:Numerical-iteration}
\end{figure}

This model solves the atomic and photonic populations at discrete sections of the fiber. This is achieved by dividing the fiber into a number of length elements $n$ as illustrated in Fig. \ref{fig:Numerical-iteration}. The rate equations are solved at each time step for each length element using the fourth order Runge-Kutta (RK4) method. Time and space are coupled such that the time step $\varDelta t$ is defined as the time required for light to traverse a single fiber element of length $\varDelta L=\frac{L}{n}$ \cite{gorjan2011role}. The photonic populations are shifted one length element at each time step. At the fiber ends, these populations are reflected from, or transmitted through, the resonator mirrors.

The model FLAPP is an enhancement of that developed by Gorjan \textit{et al.} \cite{gorjan2011role} because it solves the rate equations for a single 2D population matrix that contains all populations of all fiber elements. This results in a significant reduction in computational time. The model also uses 3D parameter matrices so that parameters may be varied for multiple parallel simulations.

\subsection{Pump absorption}

The first pump has a wavelength between 966 nm \cite{fortin2016watt} and 985 nm \cite{hendersonsapir2014midinfrared} depending on the experiment and is labeled $\mathrm{P_{1}}$. This pump launches photons into either the core \cite{hendersonsapir2014midinfrared} or the inner cladding \cite{hendersonsapir2016versatile,fortin2016watt} at one end of the fiber. The ions that absorb these pump photons are excited from energy level $\mathrm{^{4}I_{15/2}}$ (1) to level $\mathrm{^{4}I_{11/2}}$ (3) by ground state absorption (GSA). 

The second pump, $\mathrm{P_{2}}$, has a wavelength between 1973~nm \cite{hendersonsapir2014midinfrared} and 1976 nm \cite{fortin2016watt} depending on the experiment and is launched into the fiber core. The ions in level $\mathrm{^{4}I_{11/2}}$ (3) that absorb photons from the second pump are excited to level $\mathrm{^{4}F_{9/2}}$ (5). The model allows for pumping from either end of the fiber or from both ends simultaneously.

There are two excited state absorption (ESA) processes associated with the first pump - one from level $\mathrm{^{4}I_{11/2}}$ (3) to level $\mathrm{^{4}F_{7/2}}$ (7) and one from level $\mathrm{^{4}I_{9/2}}$ (4) to level $\mathrm{^{4}F_{3/2}}$ followed by fast multi-phonon decay to level $\mathrm{^{4}F_{7/2}}$ (7). The latter excited state absorption is treated as a direct transition from levels $\mathrm{^{4}I_{9/2}}$ (4) to $\mathrm{^{4}F_{7/2}}$ (7) due to the fast multi-phonon decay.

\subsection{Relaxation}

Each excited energy level has an intrinsic lifetime, $\tau$, and relaxation rate, $r=\tau^{-1}$ which includes radiative (fluorescence) and non-radiative multi-phonon (MP) decay. Relaxation from an upper level $i$ to a lower level $j$ has an associated branching ratio $\beta_{ij}$ where $\sum_{j=1}^{i-1}\beta_{ij}=1$. Then the relaxation rate $r_{ij}$ from an upper level $i$ to lower level $j$ is given by $r_{ij}=\beta_{ij}r_{i}=\beta_{ij}\tau_{i}^{-1}$.

\subsection{Lasing \label{sub:Lasing}}

Pump absorption of $\mathrm{P_{1}}$ and $\mathrm{P_{2}}$ photons creates a population inversion between the $\mathrm{^{4}I_{9/2}}$ (4) and $\mathrm{^{4}F_{9/2}}$ (5) energy levels. Spontaneous emission (radiative decay) from level $\mathrm{^{4}F_{9/2}}$ (5) to level $\mathrm{^{4}I_{9/2}}$ (4) initiates lasing at a wavelength of 3.5~\textmu m. Similarly, radiative decay from level $\mathrm{^{4}I_{11/2}}$ (3) to level $\mathrm{^{4}I_{13/2}}$ (2) can initiate lasing at a wavelength of 2.8~\textmu m.

The rate of spontaneous emission $R_{sp}$ is proportional to the population of the upper laser level $N_{upper}$. The rate of stimulated emission $R_{se}$ is proportional to the laser photon density $F_{laser}$  multiplied by $N_{upper}$. Note that laser photons may be reabsorbed by ions in the lower laser level $N_{lower}$, reducing the effective gain of stimulated emission. The rate of absorption is proportional to $F_{laser}$ and $N_{lower}$. The cross sections of these transitions are elaborated in Section \ref{sec:Cross-sections}.

\subsection{Interionic processes}

Energy transfer processes are non-radiative energy exchanges that occur between ions. The significance of these interactions is often dependent on doping concentration \cite{golding2000energy,li2012numerical} which determines the mean spacing between ions and hence the probability of interactions.

The relevant energy transfer process $W_{ijkl}$ describes the non-radiative energy exchange between two ions initially in levels $i$ and $j$ that transition to levels $k$ and $l$. The energy transfer processes in Er\textsuperscript{3+}-doped ZBLAN that have been published in literature are $W_{2214}$, $W_{3317}$, $W_{6142}$, and $W_{5362}$ \cite{golding2000energy,hendersonsapir2016new}. The rate $W_{ijkl}$ is expressed in units of volume per unit time.

\subsection{Rate equations\label{sec:Rate-equations}}

The atomic and photonic rate equations are stated below with spatial and time dependence omitted for brevity. A detailed explanation of each term follows.

The atomic rate equations are:

\begin{align}
\frac{dN_{7}}{dt}= & -\sum_{i=1}^{6}r_{7i}N_{7}+R_{P_{1}abs_{37}}F_{P_1}+R_{P_{1}abs_{47}}F_{P_1}\nonumber \\
&+W_{3317}N_{3}^{2}\\
\frac{dN_{6}}{dt}= & r_{76}N_{7}-\sum_{i=1}^{5}r_{6i}N_{6}-W_{6142}N_{6}N_{1}+W_{5362}N_{5}N_{3}\\
\frac{dN_{5}}{dt}= & \sum_{i=6}^{7}r_{i5}N_{i}-\sum_{i=1}^{4}r_{5i}N_{5}+R_{P_{2}abs_{35}}F_{P_2}\nonumber \\
&-R_{L_{2}se_{54}}F_{L_{2}}-W_{5362}N_{5}N_{3}\\
\frac{dN_{4}}{dt}= & \sum_{i=5}^{7}r_{i4}N_{i}-\sum_{i=1}^{3}r_{4i}N_{4}-R_{P_{1}abs_{47}}F_{P_1}\nonumber \\
&+R_{L_{2}se_{54}}F_{L_{2}}+W_{2214}N_{2}^{2}+W_{6142}N_{6}N_{1}\\
\frac{dN_{3}}{dt}= & \sum_{i=4}^{7}r_{i3}N_{i}-\sum_{i=1}^{2}r_{3i}N_{3}+R_{P_{1}abs_{13}}F_{P_1}\nonumber \\
&-R_{P_{1}abs_{37}}F_{P_1}-R_{P_{2}abs_{35}}F_{P_2}\nonumber \\
&-R_{L_{1}se_{32}}F_{L_{1}}-2W_{3317}N_{3}^{2}-W_{5362}N_{5}N_{3}\\
\frac{dN_{2}}{dt}= & \sum_{i=3}^{7}r_{i2}N_{i}-r_{21}N_{2}+R_{L_{1}se_{32}}F_{L_{1}}-2W_{2214}N_{2}^{2}\nonumber \\
&+W_{6142}N_{6}N_{1}+W_{5362}N_{5}N_{3}\\
\frac{dN_{1}}{dt}= & \sum_{i=2}^{7}r_{i1}N_{i}-R_{P_{1}abs_{13}}F_{P_{1}}+W_{2214}N_{2}^{2}\nonumber \\
&+W_{3317}N_{3}^{2}-W_{6142}N_{6}N_{1}\\
N_{core}= & \sum_{i=1}^{7}N_{i}
\end{align}

where $N_{i}$ is the population density of ions in energy level $i$, $N_{core}$ is the doping density of ions in the fiber core, $F$ is the photonic population number density, and $r_{ij}$ is the intrinsic relaxation rate from level $i$ to $j$. $R_{P_{k}abs_{ij}}$ is the rate of pump absorption of pump $\mathrm{P_k}$ with transitions from level $i$ to $j$ and $R_{L_{k}se_{ij}}$ is the rate of stimulated emission of laser $\mathrm{L_k}$ with transition from level $i$ to $j$. $W_{ikjl}$ is the rate of interionic interactions resulting in transitions from levels $i$ and $j$ to $k$ and $l$.

The absorption and stimulated emission rates include the populations of the upper and lower laser energy levels and the cross sections of transitions between them.

\begin{align}
R_{P_{k}abs_{ij}}&=v\left(\sigma_{P_{k}abs_{ij}}N_{i}-\sigma_{P_{k}em_{ji}}N_{j}\right)\label{eq:Rabs}\\
R_{L_{k}se_{ij}}&=v\sigma_{L_{k}se_{ij}}\left(b_{i}N_{i}-\frac{g_{i}}{g_{j}}b_{j}N_{j}\right)\\
&=v\left(\sigma_{L_{k}em_{ij}}N_{i}-\sigma_{L_{k}abs_{ji}}N_{j}\right)\label{eq:Rse}
\end{align}

where $v=\frac{c_{0}}{n_{core}}$ is the speed of light inside the fiber, approximated to be constant for pump and laser wavelengths. $\sigma_{P_{k}abs_{ij}}$ and $\sigma_{P_{k}em_{ji}}$ are the effective cross sections of absorption and emission of pump $\mathrm{P_k}$ for the transition between levels $i$ and $j$. $\sigma_{L_{k}se_{ij}}$ is the cross section of stimulated emission of laser $\mathrm{L_k}$ for the transition from level $i$ to $j$. $\sigma_{L_{k}em_{ij}}$ and $\sigma_{L_{k}abs_{ji}}$ are the effective cross sections of emission and absorption of laser $\mathrm{L_k}$. $b_{i}$ and $b_{j}$ are the Boltzmann factors of the upper and lower laser sublevels $i$ and $j$. $g_{i}$ and $g_{j}$ are the degeneracies of the upper and lower laser sublevels.

For Er\textsuperscript{3+} ions, each energy level is split into $\left(2J+1\right)/2$ Stark levels, where $J$ is the total angular momentum quantum number, leaving Kramers degeneracy \cite{huang2001stark}. Therefore, the Stark levels in Er\textsuperscript{3+} (having odd number of electrons) have degeneracies $g_{i}=g_{j}=2$ \cite{li2012numerical}.


Photonic population densities are calculated for propagation in each of the `$\mathrm{+}$' and `$\mathrm{-}$' directions illustrated in Fig. \ref{fig:Numerical-iteration}. Note that we calculate the photonic populations inside the core only since only these populations are available to interact with the Er\textsuperscript{3+} ions. The mode fields extend outside the core for wavelengths that have fiber $V$ parameters smaller than 2.4 (see Section \ref{sec:Mode-overlap}) such that only the single transverse mode is guided. Therefore, we include a mode overlap factor $\varGamma$ to correct for rates of change of population density within the core \cite{becker1999erbium}.

The photonic rate equations are:

\begin{align}
\frac{dF_{P_1}^{\pm}}{dt}= & \varGamma_{P_1}\left[-R_{P_{1}abs_{13}}-R_{P_{1}abs_{37}}-R_{P_{1}abs_{47}}\right]F_{P_1}^{\pm}\nonumber \\
&+R_{P_{1}loss}F_{P_1}^{\pm}\\
\frac{dF_{P_2}^{\pm}}{dt}= & \varGamma_{P_2}\left[-R_{P_{2}abs_{35}}F_{P_2}^{\pm}\right]+R_{P_{2}loss}F_{P_2}^{\pm}\\
\frac{dF_{L_{1}}^{\pm}}{dt}= & \varGamma_{L_{1}}\left[R_{L_{1}se_{32}}F_{L_{1}}^{\pm}+R_{L_{1}sp_{32}}\right]+R_{L_{1}loss}F_{L_{1}}^{\pm}\label{eq:dFdt2800}\\
\frac{dF_{L_{2}}^{\pm}}{dt}= & \varGamma_{L_{2}}\left[R_{L_{2}se_{54}}F_{L_{2}}^{\pm}+R_{{L_2}sp_{54}}\right]+R_{L_{2}loss}F_{L_{2}}^{\pm}\label{eq:dFdt3470}
\end{align}

where $F^{\pm}$ is the density of pump $\mathrm{P_k}$ of laser $\mathrm{L_k}$ photons propagating in the `$\pm$' direction and $\varGamma$ is the mode overlap factor of the photon field as defined in Section \ref{sec:Mode-overlap}. $R_{L_{k}loss}$ and $R_{P_{k}loss}$ are the loss rates of laser and pump photons. $R_{L_{k}sp_{ij}}$ is the spontaneous emission rate of laser $\mathrm{L_k}$ photons from level $i$ to $j$.

The rate term for spontaneous emission is given by

\begin{equation}
R_{L_{k}sp_{ij}}=\frac{f_{accept}}{2}f_{rad}r_{ij}N_{i}
\end{equation}

where $f_{accept}$ is the probability of acceptance of a spontaneously emitted photon being trapped in the fiber in either direction of propagation, and $f_{rad}$ is the probability that the relaxation is radiative.

\subsection{Cross sections\label{sec:Cross-sections}}

For two nondegenerate states of manifold sublevels $i$ and $j$, the emission and absorption cross sections will be equal, i.e., $\sigma_{ij}=\sigma_{ji}$ \cite{siegman1986lasers}. In the case of rare earth ions doped into a host, any `level' is actually a manifold of sublevels and a transition of energy separation $h\nu$ can occur between multiple pairs of sublevels. The energy of each sublevel is slightly dependent on the host because variations in the local electric fields cause Stark shifts. The inhomogeneous nature of glass hosts means that effective sublevel energy positions are blurred rather than discrete. 

The populations of each sublevel is dependent on thermal distribution and can be estimated by multiplication of the total level population by the Boltzmann factor of the sublevel. Therefore, a cross section must be defined at a particular frequency within the spectral bandwidth of the transition. A cross section derived	from experimental measurement of absorption or emission spectra is an \textit{effective} cross section, i.e., inclusive of the sublevel Boltzmann factor, at a given temperature.

Measured energy level positions and Stark splitting of Er\textsuperscript{3+} at 13 K are provided by Huang \textit{et al.} \cite{huang2001stark} in units of cm\textsuperscript{-1}. We use this data and the McCumber relation \cite{mccumber1964einstein} to calculate effective emission cross sections from absorption cross sections and vice versa. The McCumber relation considers that the population of each sublevel is determined by the Boltzmann population distribution.

\subsection{Mode overlap}\label{sec:Mode-overlap}

A Gaussian intensity profile is a good approximation for a single mode field inside a step index fiber \cite{becker1999erbium}. The intensity of each mode has a central peak and tends to zero away from longitudinal axis of the fiber. We define the mode overlap $\varGamma$ at any given wavelength as the fraction of the power in the mode that overlaps the core. Only the overlapped portion of the mode is available to interact with the Er\textsuperscript{3+} ions. The entire mode, however, will be affected by gain or loss as a result of these interactions.

To estimate the mode overlap $\varGamma$ of single mode operation, we take the approach of calculating the fraction of power transmitted through a circular aperture of radius equal to the fiber core radius $a$. This applies to the case of a lowest order mode only, which is well approximated by a Gaussian beam.

The intensity $I$ of a Gaussian beam inside the fiber at a distance $r$ from the longitudinal axis may be derived from the equation for the complex wave amplitude $\tilde{u}\left(x,y,z\right)$ given by Siegman \cite{siegman1986lasers}. Intensity is given by $I(r)=I_{0}\exp\left(-2\frac{r^{2}}{w^{2}}\right)$ where $w$ is the mode field radius. Then, using integration by substitution, we can calculate the mode overlap factor $\varGamma$ as the ratio of power inside the core $P_{core}$ of radius $a$ to total power in the mode $P_{mode}$.

\begin{equation}
\varGamma=\frac{P_{core}}{P_{mode}}=\frac{\int_{0}^{2\pi}\int_{0}^{a}I(r)rdrd\phi}{\int_{0}^{2\pi}\int_{0}^{\infty}I(r)rdrd\phi}=1-\exp\left(-2\frac{a^{2}}{w^{2}}\right)\label{eq:Mode overlap}
\end{equation}

The mode field radius (or spot size) $w$ for step index, single-mode fibers is estimated by the Marcuse empirical formula \cite{marcuse1977loss}:

\begin{equation}
w\approx a\left(0.65+\frac{1.619}{V^{3/2}}+\frac{2.879}{V^{6}}\right)\label{eq:Mode field radius}
\end{equation}

where $V$ is a parameter for step-index fibers defined by $V=\frac{2\pi}{\lambda}a\left(NA\right)\label{eq:V number}$ and $NA$ is the numerical aperture.

The mode overlap $\varGamma$ of a highly multi-mode beam, such as the clad pumping of a double clad fiber, is estimated to be the ratio $A_{core}/(A_{core}+A_{clad})$, where $A_{core}$ and $A_{clad}$ are the cross sectional areas of the core and clad.

The relation between the photon density inside the core $F$ to total power in the mode $P_{mode}$ is given by

\begin{equation}
F\left(\nu\right)=\frac{\bar{I}_{core}\left(\nu\right)n_{core}\left(\nu\right)}{h\nu c_{0}}=\varGamma\left(\nu\right)\frac{P_{mode}\left(\nu\right)n_{core}\left(\nu\right)}{Ah\nu c_{0}}
\end{equation}

where $\nu$ is the photon frequency, $A$ is the cross sectional area of the core, and $\bar{I}_{core}=\frac{P_{core}}{A}$ is the mean intensity inside the core in the transverse plane, i.e., the transverse intensity profile inside the core is assumed to be uniform.

Any change in photon population is distributed throughout the mode which extends beyond the fiber core. The mode overlap factor is implemented in the photonic rate equations to convert from rates of change of photon population in the mode $F_{mode}$ to rates of change of photon population inside the fiber core $F$.

\begin{equation}
\frac{dF}{dt}\approx\frac{\varDelta F}{\varDelta t}=\varGamma\frac{\varDelta F_{mode}}{\varDelta t}
\end{equation}

\subsection{Loss rate}

The loss rate is calculated from an internal loss coefficient that arises from scatter and absorption of pump and laser photons due to the glass host. This loss rate does not include transmission losses through resonator mirrors. The loss of laser photons is given by:

\begin{align}
\left(\phi\left(z\right)\right)_{loss}&=\phi_{0}\exp\left(-\alpha z\right)
\end{align}

where $\phi\left(z\right)$ is the number of photons (proportional to power) at distance $z$ along the fiber propagating in either direction, $\phi_{0}$ is the number of photons at $z=0$, and $\alpha$ is the measured \textit{internal loss coefficient}, assumed to be constant along the length of the fiber. Therefore, the loss over a single fiber element length $\varDelta L$ in time step $\varDelta t$ is

\begin{align}
\left(\frac{\varDelta\phi}{\varDelta t}\right)_{loss}&\approx\frac{\exp\left(-\alpha\varDelta L\right)-1}{\varDelta t}\phi_{0}
\end{align}

The rates of internal loss $R_{loss}$ from the photonic rate Eqns. \ref{eq:dFdt2800} and \ref{eq:dFdt3470} at each wavelength $\lambda$ are given by

\begin{align}
\left(\frac{dF}{dt}\right)_{loss}=R_{loss}&\approx\frac{\exp\left(-\alpha\varDelta L\right)-1}{\varDelta t}
\end{align}

Two types of resonator mirror losses are also implemented in the model. The first is a \textit{scattering loss} that reduces transmission without affecting reflection. The second is a \textit{reflection efficiency} that reduces the effective mirror reflectivity without affecting transmission.


\section{Simulation parameters}\label{sec:simulation-parameters}

The model, FLAPP, was tested on three experiments published in literature with their respective properties listed in Table \ref{tab:experimental_properties}. The first experiment (H2014) \cite{hendersonsapir2014midinfrared} used a single clad fiber manufactured by IR-Photonics (IRP). The second (H2016) \cite{hendersonsapir2016versatile} and third (F2016) \cite{fortin2016watt} experiments used the same design double clad fiber manufactured by Le Verre Fluor\'{e} (LVF) that had a lower doping. The first pump was launched into the inner cladding and the second pump was launched into the core. This double clad fiber has a circular inner cladding (diameter 260~\textmu m) with two parallel flats (separated by 240~\textmu m). The second experiment used discrete highly reflective (HR) and output coupler (OC) mirrors butt-coupled to the fiber whereas the third experiment used an all fiber geometry including a fiber Bragg grating (FBG), written directly into the fiber, as the output coupler.

\begin{table}[h!]
	\centering
	\caption{Experimental properties}
	\label{tab:experimental_properties}
	\begin{tabular}{TTTTT}
	\toprule
	Property & H2014 \cite{hendersonsapir2014midinfrared} & H2016 \cite{hendersonsapir2016versatile} & F2016 \cite{fortin2016watt} & Unit\\
	\midrule
	Manufacturer & IRP & LVF & LVF\\
	Cladding & single & double & double\\
	Er\textsuperscript{3+} & 1.7 & 1.0 & 1.0 & mol.\%\\
	$a$ & 5.25 & 8.25 & 8.25 & \textmu m\\
	NA(core) & 0.150 & 0.125 & 0.125 &\\
	$L$ & 0.18 & 2.80 & 4.30 & m\\
	$R_{OC}$(3.5 \textmu m)  & 90 & 80 & 55 & \%\\
	OC type & mirror & mirror & FBG &\\
	$R_{HR}$(3.5 \textmu m) & 99 & 99 & 90 & \%\\
	HR type & mirror & mirror & mirror &\\
	$P_{1}$ power & 0.194 & 2.0 & 6.5, 3.5 & W\\
	$P_{1}$ pumping & core & clad & clad &\\
	$\lambda_{P_{1}}$ & 985 & 977 & 966-974 & nm\\
	$\lambda_{P_{2}}$ & 1973 & 1973 & 1976 & nm\\
	\bottomrule
	\end{tabular}
\end{table}

Tables \ref{tab:pump_and_laser_parameters}, \ref{tab:spectroscopic_parameters}, and \ref{tab:nli_parameters} list the simulation parameters that were sourced and calculated from multiple references, including Refs. \cite{hendersonsapir2015thesis,huang2001stark,caspary2001thesis,quimby1991excited,bogdanov1999energy,gan1995optical,gorjan2011role,li2012numerical,wang2009excited}. Parameters found to be in good agreement with multiple independent sources were held fixed in the simulations. Such parameters include the absorption cross section of $\mathrm{P_1}$, stimulated emission cross section of $\mathrm{L_1}$, mirror reflectivities at each wavelength, and intrinsic lifetimes of levels $\mathrm{^{4}I_{13/2}}$ (2) and $\mathrm{^{4}I_{11/2}}$ (3). Parameters with larger uncertainties were altered independently, within their stated uncertainties, to test their significance. These parameters were likely to affect final results significantly if varied by 25\% or so from measured or published values. Such parameters include absorption cross section of pump $\mathrm{P_2}$, stimulated emission cross section of the 3.5~\textmu m laser transition, the cross relaxation parameter $W_{5362}$, and lifetimes of levels $\mathrm{^{4}I_{9/2}}$ (4) and $\mathrm{^{4}F_{9/2}}$ (5).

\begin{table}[h!]
	\centering
	\caption{Simulation parameters of pumps and lasers}
	\label{tab:pump_and_laser_parameters}
	\begin{tabular}{TTTTTT}
	\toprule
	Property & H2014  & H2016  & F2016  & Unit & Ref.\\
	& \cite{hendersonsapir2014midinfrared} & \cite{hendersonsapir2016versatile} & \cite{fortin2016watt} &\\
	\midrule
	$\lambda_{P_{1}}$ & 985 & 977 & 968 & nm\\
	$\lambda_{P_{2}}$ & 1973 & 1973 & 1976 & nm\\
	$\sigma_{P_{1}abs_{13}}$ & 9.30 & 19.5 & 8.56 & $\mathrm{10^{-26}~m^{2}}$ & \cite{quimby1991excited}\\
	$\sigma_{P_{1}abs_{37}}$ & 2.00 & 9.30 & 30.7 & $\mathrm{10^{-26}~m^{2}}$ & \cite{quimby1991excited}\\
	$\sigma_{P_{1}abs_{47}}$ & 25.5 & 13.5 & 2.10 & $\mathrm{10^{-26}~m^{2}}$ & \cite{quimby1991excited}\\
	$\sigma_{P_{2}abs_{35}}$ & 30.0 & 30.0 & 30.0 & $\mathrm{10^{-26}~m^{2}}$ & \cite{hendersonsapir2015thesis}\\
	$\sigma_{P_{1}em_{31}}$ & 11.5 & 16.1 & 4.45 & $\mathrm{10^{-26}~m^{2}}$ & \cite{quimby1991excited,huang2001stark}\\
	$\sigma_{P_{1}em_{73}}$ & 6.75 & 21.1 & 44.0 & $\mathrm{10^{-26}~m^{2}}$ & \cite{quimby1991excited,huang2001stark}\\
	$\sigma_{P_{1}em_{74}}$ & 47.8 & 17.4 & 1.70 & $\mathrm{10^{-26}~m^{2}}$ & \cite{quimby1991excited,huang2001stark}\\
	$\sigma_{P_{2}em_{53}}$ & 36.1 & 36.1 & 37.5 & $\mathrm{10^{-26}~m^{2}}$ & \cite{hendersonsapir2015thesis,huang2001stark}\\
	$\lambda_{L_{1}}$ & 2800 & 2800 & 2800 & nm\\
	$\lambda_{L_{2}}$ & 3470 & 3470 & 3440 & nm\\
	$\sigma_{L_{1}em_{32}}$ & 45.0 & 45.0 & 45.0 & $\mathrm{10^{-26}~m^{2}}$ & \cite{wang2009excited}\\
	$\sigma_{L_{2}em_{54}}$ & 12.0 & 12.0 & 10.8 & $\mathrm{10^{-26}~m^{2}}$ & \cite{hendersonsapir2015thesis,tobben1992room}\\
	$b_{2}$ & 0.210 & 0.210 & 0.210 & & \cite{huang2001stark}\\
	$b_{3}$ & 0.350 & 0.350 & 0.350 & & \cite{huang2001stark}\\
	$b_{4}$ & 0.575 & 0.575 & 0.427 & & \cite{huang2001stark}\\
	$b_{5}$ & 0.435 & 0.435 & 0.308 & & \cite{huang2001stark}\\
	$R_{OC}$(3.5 \textmu m) & 87 & 58 & 55 & \% &\\
	$R_{HR}$(3.5 \textmu m) & 99 & 99 & 86 & \% &\\
	$\alpha$(3.5 \textmu m) & 0.060 & 0.035 & 0.035 & $\mathrm{m^{-1}}$ &\\
	\bottomrule
	\end{tabular}
\end{table}

\begin{figure}[!h]
	\begin{centering}
		\includegraphics[width=0.90\columnwidth]{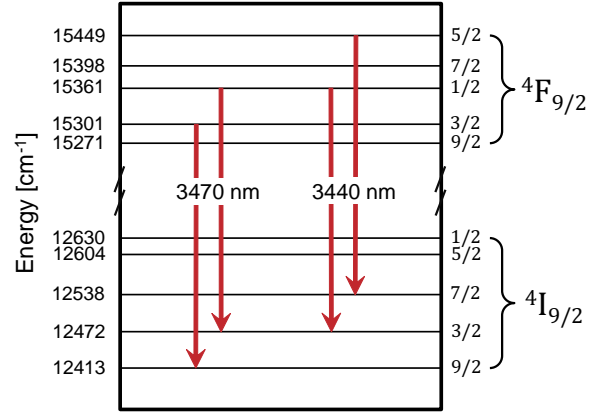}
		\caption{Stark splitting in Er\textsuperscript{3+}-doped ZBLAN and predicted laser transitions of lasers developed by Henderson-Sapir \textit{et al.} \protect\cite{hendersonsapir2014midinfrared,hendersonsapir2016versatile} (3470~nm) and Fortin \textit{et al.} \protect\cite{fortin2016watt} (3440~nm). Each arrow connects a pair of Stark levels that correspond to the predicted transitions of the associated wavelengths. The energy data was sourced from Ref. \protect\cite{huang2001stark} and the assignment order of Stark levels was sourced from Ref. \protect\cite{santos1998absorption}.}
		\label{fig:Transitions-Ori2016Laval2016}
		\par\end{centering}
\end{figure}

The emission cross sections, $\sigma_{P_{1}em}$ and $\sigma_{P_{2}em}$, of the first and second pumps were calculated from absorption measurements \cite{hendersonsapir2015thesis} using McCumber theory. The effective stimulated emission cross section $\sigma_{L_{2}em}$ of the 3.5 \textmu m laser was estimated using the 3.5 \textmu m fluorescence spectrum given by T\"{o}bben \cite{tobben1992room} and the F\"{u}chtbauer-Ladenburg equation \cite{hendersonsapir2015thesis}. The 3.5 \textmu m laser transitions, predicted from measured Stark split energy levels \cite{huang2001stark}, is shown in Fig. \ref{fig:Transitions-Ori2016Laval2016}. The Boltzmann factors $b_{i}$ of the Stark split upper and lower energy levels were calculated using the partition function $Z$ of each level as follows:

\begin{align}
b_{i}&=\frac{\exp\left(\frac{-E_{i}}{kT}\right)}{Z}=\frac{\exp\left(\frac{-E_{i}}{kT}\right)}{\sum_{i}\exp\left(\frac{-E_{i}}{kT}\right)}
\end{align}


where $k$ is the Boltzmann constant and $T$ is the temperature, assumed to be 300 K. The Boltzmann factors $b_{i}$ were then summed for each predicted laser transition for each level.

The simulation reflectivities of the butt-coupled mirrors are lower than their specified values since the light is incident from a ZBLAN fiber (refractive index $n_{core}=1.48$) rather than from air. The dielectric coatings of the mirrors used in experiments H2014 \cite{hendersonsapir2014midinfrared} and H2016 \cite{hendersonsapir2016versatile} is proprietary information and, therefore, we had to make our own predictions about their composition and number of layers. We also performed 0\textdegree\ reflectivity measurements on the 90\% and 80\% output couplers that resulted in $90\pm1$\% and $75\pm1$\% respectively. The simulation reflectivities were calibrated to give good agreement with the slope efficiencies of experimental data.

The intrinsic lifetimes $\tau_{i}$ and branching ratios $\beta_{ij}$ in Er\textsuperscript{3+}-doped ZBLAN that were used in each simulation are listed in Table \ref{tab:spectroscopic_parameters}.

\begin{table}[h!]
	\centering
	\caption{Spectroscopic parameters of Er\textsuperscript{3+}}
	\label{tab:spectroscopic_parameters}
	\begin{tabular}{TTT}
		\toprule
		Parameters & Value & Source\\
		\midrule
		$\tau_{2}$ & 9.9 ms & \cite{bogdanov1999energy}\\
		$\tau_{3}$ & 7.9 ms & \cite{bogdanov1999energy}\\
		$\tau_{4}$ & 8.0 \textmu s & \cite{bogdanov1999energy}\\
		$\tau_{5}$ & 177 \textmu s & \cite{bogdanov1999energy}\\
		$\tau_{6}$ & 530 \textmu s & \cite{bogdanov1999energy}\\
		$\tau_{7}$ & 5.0 \textmu s & \cite{pollnau2002energy}\\
		$\beta_{21}$ & 1 & \cite{bogdanov1999energy}\\
		$\beta_{32},\beta_{31}$ & 0.182, 0.818 & \cite{bogdanov1999energy}\\
		$\beta_{43},\beta_{41}$ & 0.999, 0.001 & \cite{bogdanov1999energy}\\
		$\beta_{54},\beta_{53},\beta_{52},\beta_{51}$ & 0.808, 0.008, 0.009, 0.175 & \cite{bogdanov1999energy}\\
		$\beta_{65},\beta_{64},\beta_{63},\beta_{62},\beta_{61}$ & 0.285, 0.029, 0.014, 0.193, 0.479 & \cite{bogdanov1999energy}\\
		$\beta_{76},\beta_{71}$ & 0.990, 0.010 & \cite{pollnau2002energy}\\
		\bottomrule
	\end{tabular}
\end{table}

The interionic parameters used in the simulations are listed in Table \ref{tab:nli_parameters} and are consistent with the weakly interacting regime in recent literature \cite{gorjan2011role,li2012numerical,hendersonsapir2016new}. Interionic processes in Er\textsuperscript{3+}-doped ZBLAN fibers are currently not fully understood. There are discrepancies in literature regarding the rate values of the significant processes and their dependence on doping concentration, particularly between bulk glass and fiber \cite{golding2000energy,bogdanov1999energy,gorjan2011role,li2012numerical}. ZBLAN composition and quality of fiber draw may vary between suppliers and draws. This variation may result in different distributions of Er\textsuperscript{3+}-ions \cite{srinivasan2000indirect} which affect the rates at which interionic processes occur.

\begin{table}[h!]
	\centering
	\caption{Simulation parameters of interionic processes}
	\label{tab:nli_parameters}
	\begin{tabular}{TTT}
		\toprule
		Parameters & Value $(\mathrm{10^{-24}m^{3}s^{-1}})$ & Ref.\\
		\midrule
		$W_{2214}$ & $\mathrm{0.40}$ & \cite{li2012numerical,hendersonsapir2016new}\\
		$W_{3317}$ & $\mathrm{0.08}$ & \cite{li2012numerical,hendersonsapir2016new}\\
		$W_{6142}$ & $\mathrm{0.10}$ & \cite{li2012numerical,hendersonsapir2016new}\\
		$W_{5362}$ & $\mathrm{17.0}$ & \cite{hendersonsapir2016new}\\
		\bottomrule
	\end{tabular}
\end{table}

The upper and lower energy levels of the 3.5~\textmu m laser transition are highly populated in the DWP system compared with the more common singly pumped 2.8~\textmu m Er\textsuperscript{3+}-doped ZBLAN fiber laser. The work of Bogdanov \cite{bogdanov1999energy} shows that further processes, including reverse processes, are possible. These processes could have a greater affect on the 3.5~\textmu m DWP system than the 2.8~\textmu m fiber laser. One example of such a process is $W_{3251}$ \cite{bogdanov1999energy} that would transfer ions from the virtual ground state $\mathrm{^{4}I_{11/2}}$ (3) to the upper laser level $\mathrm{^{4}F_{9/2}}$ (5). This process was not included in the model since, to the best of our knowledge, no direct measurement of its value has been made.


\section{Model validation}\label{sec:model-validation}

We identified a list of parameters that were either measured directly or published in literature and held those parameters fixed in simulations. We then varied the remaining variable parameters slightly, within their published uncertainties or our estimate of their bounds, until we found a cohesive set of parameters that gave good agreement with experimental results. These values were therefore maintained for the three systems H2014 \cite{hendersonsapir2014midinfrared}, H2016 \cite{hendersonsapir2016versatile}, and F2016 \cite{fortin2016watt}.

Steady state results gave excellent agreement between the cases where both pumps were switched on simultaneously and where $\mathrm{P_1}$ was switched on 20~ms prior to $\mathrm{P_2}$. Therefore, we switched both pumps on simultaneously for all presented simulations. The number of fiber elements $n$ chosen for each system was determined by finding the minimum $n$ such that the variation between successive simulations was negligible. These were $n_{H2014}=10$, $n_{H2016}=28$, and $n_{F2016}=43$.

\subsection{Time domain - H2014 \cite{hendersonsapir2014midinfrared}}

In this section, we study the time evolution of the atomic populations and intracavity laser power. We also show that 20~ms is sufficient time to reach steady state. In each of the following two examples, the power of pump $\mathrm{P_1}$ is held fixed at 194~mW.

\begin{figure}[!h]
	\begin{centering}
		\includegraphics[width=1.00\columnwidth]{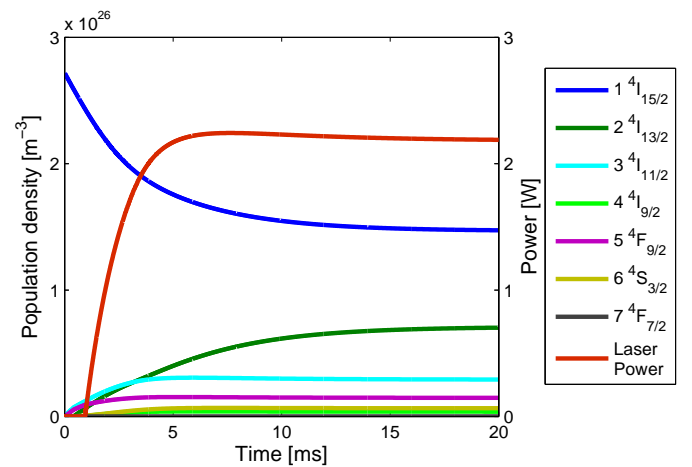}
		\caption{Modeled atomic populations of experiment H2014 by Henderson-Sapir \textit{et al.} \protect\cite{hendersonsapir2014midinfrared}. The populations are calculated midway along the fiber in the case where pump $\mathrm{P_1}$ operates at 194~mW and $\mathrm{P_2}$ operates at 2~W CW. The intracavity 3.47~\textmu m laser power propagating in the `$\mathrm{+}$' direction is plotted against the right axis.}
		\label{fig:Atomic-populations-time}
		\par\end{centering}
\end{figure}

In Fig. \ref{fig:Atomic-populations-time} the atomic populations in the middle fiber element are shown (left axis) as these approximate the mean population along the fiber. The intracavity 3.47~\textmu m laser power midway along the fiber, propagating in the `$\mathrm{+}$' direction, is also shown (right axis). Pump $\mathrm{P_2}$ operates at 2~W CW.

The threshold condition for lasing is reached at around 1~ms, after which the laser power increases rapidly over the following 4~ms. When threshold is reached, the population in level $\mathrm{^{4}F_{9/2}}$ (5) remains fairly constant beyond 1~ms. A large population is bottlenecked in level $\mathrm{^{4}I_{13/2}}$ (2) due to its relatively long intrinsic lifetime of $\tau_{2}=\mathrm{9.9\ ms}$. Therefore steady state for this system is reached at around 20~ms, making 20~ms duration simulations sufficient for steady state analysis.

\begin{figure}[!h]
	\begin{centering}
	\includegraphics[width=1\columnwidth]{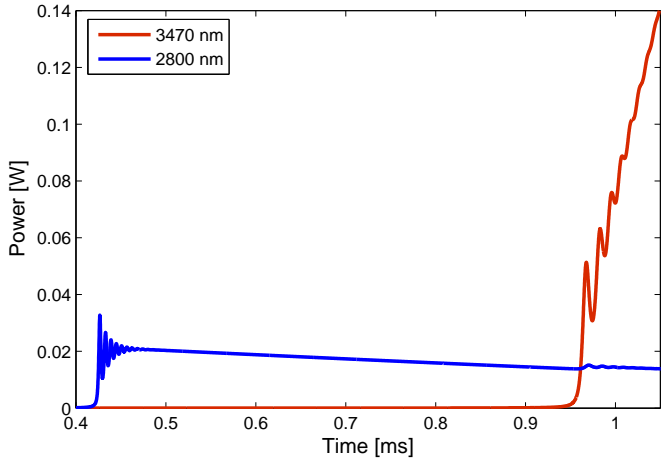}
	\par\end{centering}
	\caption{Modeled intracavity laser power of experiment H2014 by Henderson-Sapir \textit{et al.} \protect\cite{hendersonsapir2014midinfrared} immediately after threshold of the 2.8~\textmu m and 3.47~\textmu m lasers. The power is calculated midway along the fiber in the case where pump $\mathrm{P_1}$ operates at 194~mW and $\mathrm{P_2}$ operates at 2~W CW. The figure illustrates oscillatory behavior at each laser wavelength immediately after threshold. The powers plotted are for lasers propagating in the `$\mathrm{+}$' direction only.}
	\label{fig:Laser-oscillations}
\end{figure}

Intracavity laser power immediately after threshold is illustrated in Fig. \ref{fig:Laser-oscillations}. For both laser transitions, the well known relaxation oscillations \cite{siegman1986lasers} are observed. The relaxation oscillations are stronger on the 2.8~\textmu m laser emission than the 3.5~\textmu m emission due to the higher stimulated emission cross section on the 2.8~\textmu m transition. This means that 3.5~\textmu m transition relaxation oscillations have a lower frequency and damp out prior to full power being reached. The figure also illustrates that the 2.8~\textmu m laser operates at low power (20~mW intracavity power) and is gradually suppressed by the build-up of ions in its lower lasing level $\mathrm{^{4}I_{13/2}}$ (2).

\begin{figure}[!h]
	\begin{centering}
		\includegraphics[width=0.95\columnwidth]{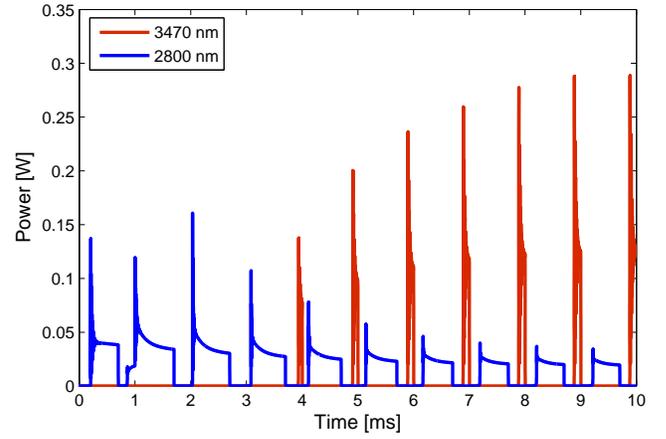}
		\caption{Modeled intracavity laser power of experiment H2014 by Henderson-Sapir \textit{et al.} \protect\cite{hendersonsapir2014midinfrared,hendersonsapir2014higher} in low power pulsed operation. The power is calculated midway along the fiber over a 10~ms simulation. Pump $\mathrm{P_2}$ operates at 200~mW with 300 \textmu s pulses at a repetition rate of 1 kHz. Lasers pulse alternately between wavelengths of 2.8~\textmu m and 3.47~\textmu m.}
		\label{fig:Laser-transmission-pulse-200mW}
		\par\end{centering}
\end{figure}

Lasing occurs at 2.8~\textmu m on the $\mathrm{^{4}I_{11/2}\rightarrow^{4}I_{13/2}}$ transition when pump $\mathrm{P_1}$ is fixed at 194~mW and the power of $\mathrm{P_2}$ is low. An interesting experimental observation occurs whilst pulsing $\mathrm{P_2}$ at low power in which lasers pulse alternately between wavelengths of 3.47~\textmu m and 2.8~\textmu m. Our simulations have reproduced this behavior as illustrated in Fig. \ref{fig:Laser-transmission-pulse-200mW}. This figure shows the modeled intracavity laser power when pump $\mathrm{P_2}$ is pulsed at low power (200~mW peak) with 300 \textmu s pulses at a repetition rate of 1 kHz. For this phenomenon to be observed, the power of $\mathrm{P_2}$ needs to be low enough to retain a sufficient population in the $\mathrm{^{4}I_{11/2}}$ (3) level between pulses to enable lasing at 2.8~\textmu m. 

\subsection{Laser output power - H2014 \cite{hendersonsapir2014midinfrared}}

The model calculates pump and laser transmission in both the `$\mathrm{+}$' and `$\mathrm{-}$' directions since both resonator mirrors are partially transmissive at each of the pump and laser wavelengths. In all of our simulations, the pumps are launched into the `$\mathrm{-}$' end of the fiber and the output coupler is located at the `$\mathrm{+}$' end of the fiber. Therefore, all transmission results that follow are transmissions in the `$\mathrm{+}$' direction.

\begin{figure}[!h]
	\begin{centering}
		\includegraphics[width=0.95\columnwidth]{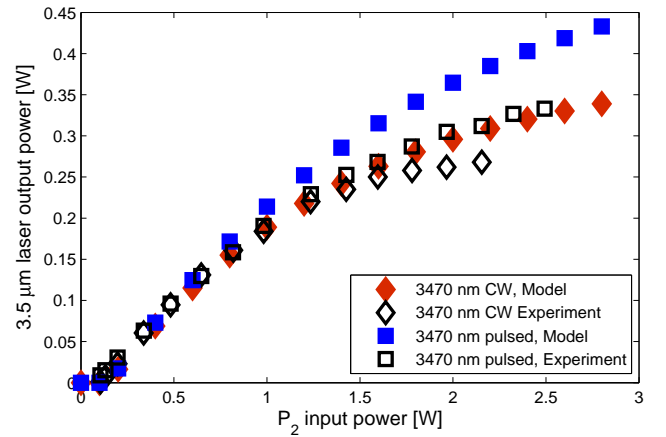}
		\caption{Modeled 3.47~\textmu m laser output power of experiment H2014 by Henderson-Sapir \textit{et al.} \protect\cite{hendersonsapir2014midinfrared} for CW and pulsed operation.}
		\label{fig:Laser-output-power-Ori2014}
		\par\end{centering}
\end{figure}

Modeled 3.47~\textmu m laser output power of experiment H2014 as a function of incident $\mathrm{P_2}$ power is presented in Fig. \ref{fig:Laser-output-power-Ori2014}. The plot shows reasonable agreement between modeled and experimental data for both CW and pulsed operation \cite{hendersonsapir2014higher}. The modeled threshold power matches experiment well at just above 100~mW. Modeled slope efficiencies and the non-linear power saturations are also closely matched to experiment.

When pulsed, the 1973~nm pump operated at a frequency of 1~kHz and a duty cycle of 30\%. The experimental pulse power was determined by dividing the average transmitted power (detected by a slow thermopile) by the duty cycle of the pulse (0.3 in this case).

The model predicts a higher saturation level than seen in experiment. This may be explained by a slight misalignment of the fiber that develops at the pump input end due to thermal expansion against the butt-coupled HR mirror. The fiber tip, initially heated by $\mathrm{P_1}$ core pumping, expands further with increasing $\mathrm{P_2}$ power since scattered pump light that is not launched into the core is absorbed by the fiber coating. The misalignment results in saturation of laser power.

\subsection{Pump transmission - H2014 \cite{hendersonsapir2014midinfrared}}

\begin{figure}[!h]
	\begin{centering}
	\includegraphics[width=0.95\columnwidth]{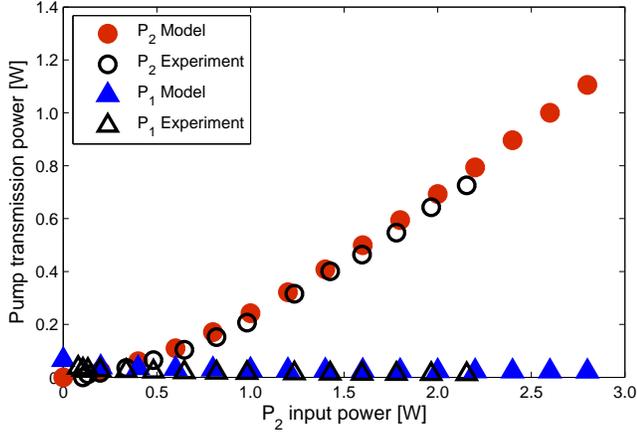}
	\par\end{centering}
	\begin{centering}
	\caption{Modeled CW pump transmission of experiment H2014 by Henderson-Sapir \textit{et al.} \protect\cite{hendersonsapir2014midinfrared} for the 985~nm ($\mathrm{P_1}$) and 1973~nm ($\mathrm{P_2}$) pumps. The incident power of $\mathrm{P_1}$ was held fixed at 194~mW while the incident power of $\mathrm{P_2}$ was varied.}
	\label{fig:Pump-transmission-Ori2014-cw}
	\par\end{centering}
\end{figure}

Modeled CW pump transmission of experiment H2014 is presented in Fig. \ref{fig:Pump-transmission-Ori2014-cw}. Transmission of both pumps $\mathrm{P_1}$ and $\mathrm{P_2}$ are shown. The plot shows good agreement between modeled and experimental data for both pumps. Nearly all of the $\mathrm{P_1}$ power is absorbed before reaching the output coupler. The transmission of pump $\mathrm{P_2}$ is dependent upon populations in levels $\mathrm{^{4}I_{11/2}}$ (3) and $\mathrm{^{4}F_{9/2}}$ (5) as well as pump absorption and emission cross sections (see Eqn. \ref{eq:Rabs}). The calculated populations in levels $\mathrm{^{4}I_{11/2}}$ (3) and $\mathrm{^{4}F_{9/2}}$ (5) are dependent on parameters that have considerable uncertainties including launch efficiency, the cross relaxation rate $W_{5362}$, and the lifetimes of the 3.5~\textmu m laser levels $\mathrm{^{4}F_{9/2}}$ (5) and $\mathrm{^{4}I_{9/2}}$ (4).

\subsection{Laser output power - H2016 \cite{hendersonsapir2016versatile} and F2016 \cite{fortin2016watt}}

\begin{figure}[!h]
	\begin{centering}
	\includegraphics[width=0.95\columnwidth]{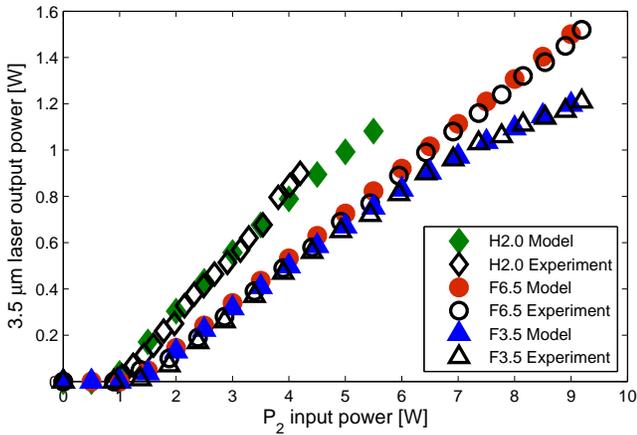}
	\caption{Modeled 3.5~\textmu m laser output powers of experiments H2016 by Henderson-Sapir \textit{et al.} \protect\cite{hendersonsapir2016versatile} and F2016 by Fortin \textit{et al.} \protect\cite{fortin2016watt}. `H2.0' refers to the experiment H2016 in which $\mathrm{P_1}$ operated at 2.0 W. `F6.5' and `F3.5' refer to experiments F2016 in which $\mathrm{P_1}$ operated at 6.5 W and 3.5 W respectively.}
	\label{fig:Laser-output-Ori2016Laval2016}
	\par\end{centering}
\end{figure}

Modeled 3.5~\textmu m laser output powers of experiments H2016 and F2016 are presented in Fig. \ref{fig:Laser-output-Ori2016Laval2016}. The plot shows good agreement between modeled and experimental data for each simulation. In each simulation, the power of $\mathrm{P_1}$ is held fixed and the power of $\mathrm{P_2}$ is incremented by 500~mW from 0~W to near the maximum power used in experiment. 

The H2016 experiment is plotted against incident pump power. The modeled launch efficiencies of pumps $\mathrm{P_1}$ and $\mathrm{P_2}$ were 90\% and 86\% respectively. The F2016 experiments are plotted against launched pump power by setting the launch efficiency of $\mathrm{P_1}$ to 100\%. However, the launch efficiency of $\mathrm{P_2}$ was set to 72\% to match the slope efficiency of experimental data. The modeled wavelength of $\mathrm{P_1}$ in experiment F2016 was 968~nm which gave good agreement with the power saturation of `F3.5'.


\section{Discussion}

\subsection{Experiment H2014 by Henderson-Sapir \textit{et al.} \cite{hendersonsapir2014midinfrared}}

\begin{figure}[!h]
	\begin{centering}
	\includegraphics[width=1\columnwidth]{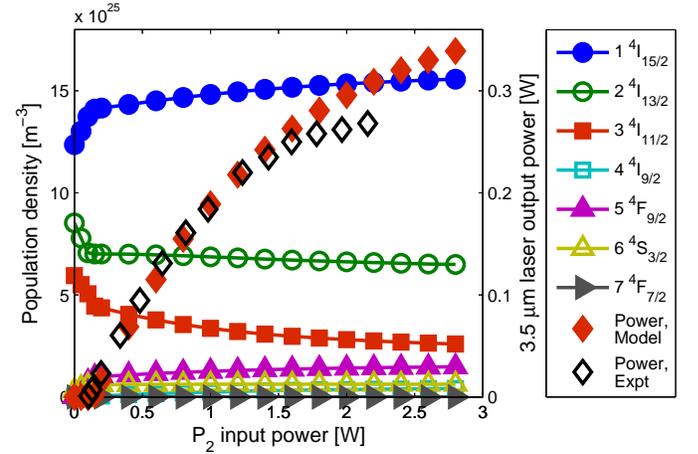}
	\par\end{centering}
	\caption{Modeled populations of experiment H2014 by Henderson-Sapir \textit{et al.} \protect\cite{hendersonsapir2014midinfrared} as a function of $\mathrm{P_2}$ pump power. The populations are averaged over the length of the fiber at the end of a 20~ms simulation. The 3.47~\textmu m laser output power is plotted against the right axis to show the relation between laser output power and steady state populations.}
	\label{fig:Final-mean-atomic-Ori2014}
\end{figure}

Modeled populations as functions of incident $\mathrm{P_2}$ power are presented in Fig. \ref{fig:Final-mean-atomic-Ori2014}. These populations are averaged over the length of the fiber once steady state has been reached. The 3.47~\textmu m laser output power is also overplotted against the right axis to show the relation between laser output power and steady state populations.

Below threshold, the population of the $\mathrm{^{4}I_{11/2}}$ (3) level decreases significantly with increasing $\mathrm{P_2}$ power as its population is pumped to the upper laser level $\mathrm{^{4}F_{9/2}}$ (5). This decrease is closely followed by a decrease in the population of level $\mathrm{^{4}I_{13/2}}$ (2), as relaxation from the depleting $\mathrm{^{4}I_{11/2}}$ (3) level is reduced. The population of the ground state increases as ions are effectively returned to the ground state. This is most likely due to ions in the now populated upper laser level $\mathrm{^{4}F_{9/2}}$ (5) returning to the ground state by radiative decay $r_{51}=989\ \mathrm{s^{-1}}$. Another likely path is cross relaxation $W_{5362}$ up to level $\mathrm{^{4}S_{3/2}}$ (6) followed by decay to the ground state, either directly ($r_{61}=904\ \mathrm{s^{-1}}$) or via level $\mathrm{^{4}I_{13/2}}$ (2) ($r_{62}=364\ \mathrm{s^{-1}}$). This cross relaxation process becomes significant as the population of the $\mathrm{^{4}F_{9/2}}$ (5) level increases.

Once threshold is achieved, the population of the $\mathrm{^{4}F_{9/2}}$ (5) level is almost clamped due to gain saturation. Perfect clamping is not achieved due to the accumulation of ions in the lower laser level $\mathrm{^{4}I_{9/2}}$ (4). Further increases in the rate of stimulated emission cause this lower lasing state population to gradually increase. This forces the population in level $\mathrm{^{4}F_{9/2}}$ (5) to slightly increase so that threshold round trip gain is maintained. Lasing causes a significant increase in the transfer rate of ions from the $\mathrm{^{4}F_{9/2}}$ (4) state to the $\mathrm{^{4}I_{11/2}}$ (3) state, thereby limiting the effect on depopulation of this level by $\mathrm{P_2}$ pumping.

The non-linear behavior seen in Figs. \ref{fig:Laser-output-power-Ori2014} and \ref{fig:Final-mean-atomic-Ori2014} at $\mathrm{P_2}$ powers greater than 1~W can be explained by power saturation due to the limited supply of ions in level $\mathrm{^{4}I_{11/2}}$ (3) available for pumping to the upper laser level $\mathrm{^{4}F_{9/2}}$ (5). Further evidence of this is the increase in relative $\mathrm{P_2}$ transmission seen in Fig. \ref{fig:Pump-transmission-Ori2014-cw} above 1~W.  

The gain medium becomes transparent to the $\mathrm{P_2}$ pump when the population of level $\mathrm{^{4}I_{11/2}}$ (3) is 20\% higher than the population of level $\mathrm{^{4}F_{9/2}}$ (5). This is calculated based on the effective $\mathrm{P_2}$ emission cross section that was calculated from the effective pump absorption cross section using McCumber theory. In Fig. \ref{fig:Final-mean-atomic-Ori2014}, the ratio of populations in levels $\mathrm{^{4}I_{11/2}}$ (3) to $\mathrm{^{4}F_{9/2}}$ (5) is 1.75 when the incident $\mathrm{P_2}$ power is 2.8~W. This occurs because the population of the $\mathrm{^{4}I_{11/2}}$ (3) level decreases and the population of the $\mathrm{^{4}F_{9/2}}$ (5) level increases due to bottlenecking of population in the  $\mathrm{^{4}I_{9/2}}$ (4) as mentioned earlier. The population of the $\mathrm{^{4}I_{11/2}}$ (3) level is reduced in two ways. Firstly, more ions are stored in the two levels above it. Secondly, the increase in level $\mathrm{^{4}F_{9/2}}$ (5) increases the number of ions that escape the cycle between the virtual ground state $\mathrm{^{4}I_{11/2}}$ (3) and upper laser level $\mathrm{^{4}F_{9/2}}$ (5). This is because the spontaneous emission rate from the $\mathrm{^{4}F_{9/2}}$ (5) level to the ground state is nine times higher than decay from the $\mathrm{^{4}I_{11/2}}$ (3) level. The cross relaxation process $W_{5362}$ increases this further.

\subsection{Parameter significance}

\begin{figure}[!h]
	\begin{centering}
	\includegraphics[width=0.95\columnwidth]{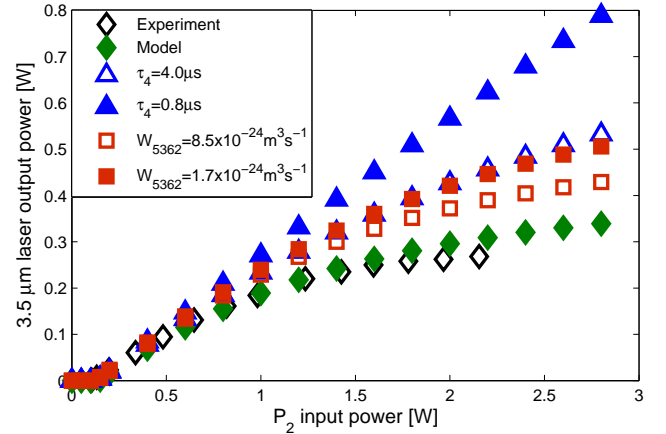}
	\par\end{centering}
	\caption{Modeled CW 3.47~\textmu m laser output powers as functions of $\mathrm{P_2}$ pump power based on variations of experiment H2014 by Henderson-Sapir \textit{et al.} \protect\cite{hendersonsapir2014midinfrared}. The intrinsic lifetime $\tau_{4}$ of the lower laser level $\mathrm{^{4}I_{9/2}}$ (4) and the cross relaxation parameter $W_{5362}$ are reduced by factors of 2 and 10 separately.}
	\label{fig:Parameter significance - t4CR}
\end{figure}

In this section we investigate the impact of changes in the lower laser level $\mathrm{^{4}I_{9/2}}$ (4) lifetime and the energy exchange process $W_{5362}$. The benchmark values are those listed in Tables \ref{tab:spectroscopic_parameters} and \ref{tab:nli_parameters}. The steady state 3.47~\textmu m output power as a function of incident $\mathrm{P_2}$ pump power is presented in Fig. \ref{fig:Parameter significance - t4CR} for a variety of lower laser state lifetimes and $W_{5362}$ values. In each case, the fiber is pumped by CW pump sources. 


The intrinsic lifetime of the lower laser level is reduced by factors of two and ten from the benchmark value $\tau_{4}=8.0$~\textmu s. The results clearly show that a dominant limitation on laser performance is the bottlenecking of ions in the lower laser level $\mathrm{^{4}I_{9/2}}$ (4) since the power saturation is removed when the lifetime of this state is reduced by a factor of 10. Ions that accumulate in this lower laser level $\mathrm{^{4}I_{9/2}}$ (4) are delayed in their return to the virtual ground state $\mathrm{^{4}I_{11/2}}$ (3) and limit the potential rate of $\mathrm{P_2}$ absorption. The population of ions in level $\mathrm{^{4}I_{9/2}}$ (4) are available to absorb photons of the 3.5 \textmu m transition and reduce the net rate of laser photon generation.

The cross relaxation parameter $W_{5362}$ is reduced by factors of two and ten from the benchmark value $W_{5362}=17.0\times10^{-24}~\mathrm{m^{3}s^{-1}}$ \cite{hendersonsapir2016new}. The rate at which these interionic interactions occur is proportional to the populations in levels $\mathrm{^{4}I_{11/2}}$ (3) and $\mathrm{^{4}F_{9/2}}$ (5). This interionic process limits the laser performance by depleting ions from the virtual ground level $\mathrm{^{4}I_{11/2}}$ (3) as well as the upper laser level $\mathrm{^{4}F_{9/2}}$ (5) which reduces stimulated emission and pump absorption. The plot of the reduced $W_{5362}=1.7\times10^{-24}~\mathrm{m^{3}s^{-1}}$ illustrates the negative effect this parameter has on laser performance by reducing slope efficiency and increasing power saturation. It is worth noting that significantly reducing this parameter does not remove the power saturation completely as this is dominated by the lower lasing state lifetime. The effect of this energy transfer also reduces significantly when double clad fibers and lower doping densities are used as described below.

\begin{figure}[!h]
	\begin{centering}
	\includegraphics[width=0.95\columnwidth]{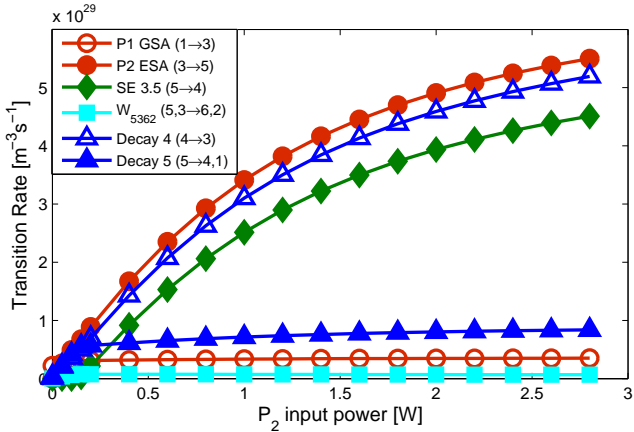}
	\par\end{centering}
	\caption{Modeled transition rates of experiment H2014 by Henderson-Sapir \textit{et al.} \protect\cite{hendersonsapir2014midinfrared}. Rates are averaged over the length of the fiber and plotted as functions of incident $\mathrm{P_2}$ pump power. The transitions are pump absorption of $\mathrm{P_1}$ and $\mathrm{P_2}$ (P1 GSA and P2 ESA), stimulated emission (SE) of the 3.5~\textmu m laser, the cross relaxation process $W_{5362}$, and the decay rates from the lower (4) and upper (5) laser levels.}
	\label{fig:Transition rates - Ori2014}
\end{figure}

To understand what processes are important in governing the performance of the DWP laser, we calculated the transition rates as functions of incident $\mathrm{P_2}$ pump power. The most significant transition rates are averaged over the length of the fiber and plotted in Fig. \ref{fig:Transition rates - Ori2014}. The most significant transition is from levels $\mathrm{^{4}I_{11/2}}$ (3) to $\mathrm{^{4}F_{9/2}}$ (5) by $\mathrm{P_2}$ pump absorption. The non-linearity of this pump absorption, as well as stimulated emission, illustrates power saturation due to depletion of ions in level $\mathrm{^{4}I_{11/2}}$ (3) as discussed earlier. The population of the lower laser level $\mathrm{^{4}I_{9/2}}$ (4) is fed mainly by stimulated emission and non-radiative transitions from the $\mathrm{^{4}F_{9/2}}$ (5) level and hence the rate of decay from this state grows with the power of the second pump.

\begin{figure}[!h]
	\begin{centering}
	\includegraphics[width=0.95\columnwidth]{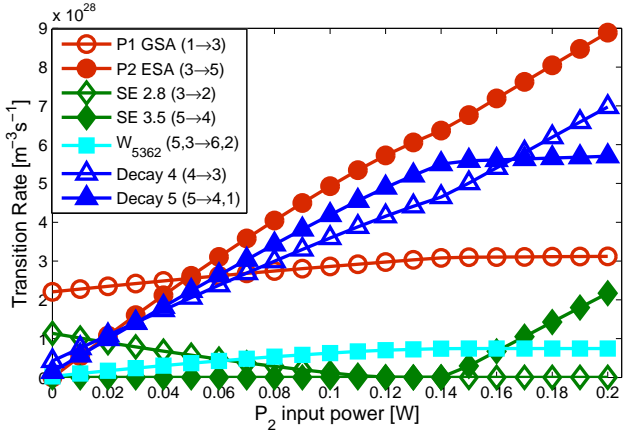}
	\par\end{centering}
	\caption{Modeled transition rates of experiment H2014 by Henderson-Sapir \textit{et al.} \protect\cite{hendersonsapir2014midinfrared} around threshold. The rates are averaged over the length of the fiber and plotted as functions of incident $\mathrm{P_2}$ pump power. The transitions are pump absorption $\mathrm{P_1}$ and $\mathrm{P_2}$ (P1 GSA and P2 ESA), stimulated emission (SE) of the 2.8~\textmu m and 3.5~\textmu m lasers, the cross relaxation process $W_{5362}$, and the decay rates of the lower (4) and upper (5) laser levels.}
	\label{fig:Transition rates around threshold - Ori2014}
\end{figure}

The variations in the transition rates as functions of incident $\mathrm{P_2}$ pump power around threshold are plotted in Fig. \ref{fig:Transition rates around threshold - Ori2014}. Below threshold, approximately 18\% of ions in the upper laser level $\mathrm{^{4}F_{9/2}}$ (5) decay to the ground state, exiting the second pump cycle. Beyond threshold of the 3.47~\textmu m laser, stimulated emission increases sharply and causes a faster return of ions to the $\mathrm{^{4}I_{11/2}}$ (3) level.

At $\mathrm{P_2}$ power levels below 70~mW the rate of stimulated emission due to the 2.8~\textmu m laser transition is more significant than $W_{5362}$. Above 70~mW, the 2.8~\textmu m laser is suppressed by absorption of the $\mathrm{P_2}$ pump and its subsequent reduction of the $\mathrm{^{4}I_{11/2}}$ (3) population. The rate in which ions leave level  $\mathrm{^{4}F_{9/2}}$ (5) by the energy transfer process $W_{5362}$ rises until threshold is reached at which point it flattens considerably. The rate in which ions are excited by $\mathrm{P_2}$ shows signs of initial saturation until the threshold of the 3.47~\textmu m laser is reached and then it resumes its linear increase.

\subsection{Experiments H2016 by Henderson-Sapir \textit{et al.} \cite{hendersonsapir2016versatile} and F2016 by Fortin \textit{et al.}\cite{fortin2016watt}}

The simulations F6.5 and F3.5 presented in Fig. \ref{fig:Laser-output-Ori2016Laval2016} are particularly sensitive to the wavelength of $\mathrm{P_1}$ due to steep variation in ground state absorption between 965~nm and 972~nm \cite{quimby1991excited}. The wavelength 968~nm provided the best fit to experimental data. This wavelength also corresponds to the peak of the excited state absorption for the $\mathrm{^{4}I_{11/2}} \rightarrow \mathrm{^{4}F_{7/2}}$ transition. The rate of excited state absorption in the H2.0 simulation is more sensitive to variations in wavelength since at 977~nm the change in cross section with wavelength is significant.

\begin{figure}[!h]
	\begin{centering}
	\includegraphics[width=0.95\columnwidth]{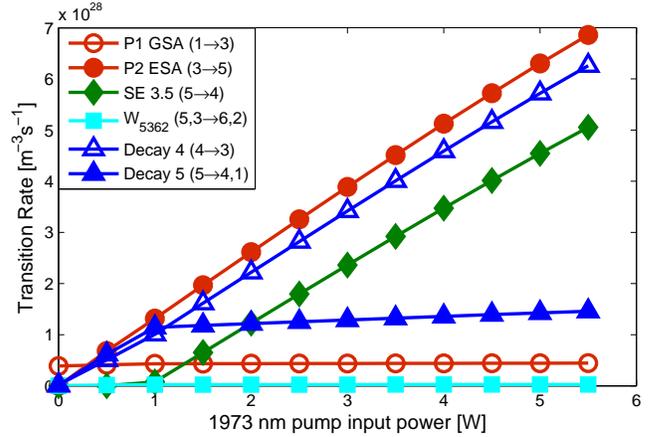}
	\par\end{centering}
	\caption{Modeled transition rates of experiment H2016 by Henderson-Sapir \textit{et al.} \protect\cite{hendersonsapir2016versatile}. The rates are averaged over the length of the fiber and plotted as functions of incident $\mathrm{P_2}$ pump power. The transitions are pump absorption of $\mathrm{P_1}$ and $\mathrm{P_2}$ (P1 GSA and P2 ESA), stimulated emission (SE) of the 3.5~\textmu m laser, the cross relaxation process $W_{5362}$, and the decay rates of the lower (4) and upper (5) laser levels.}
	\label{fig:Transition rates - Ori2016}
\end{figure}

Modeled transition rates of experiment H2016 \cite{hendersonsapir2016versatile} as functions of incident $\mathrm{P_2}$ pump power is plotted in Fig. \ref{fig:Transition rates - Ori2016}. The rates of stimulated emission and $\mathrm{P_2}$ absorption are closer to linearity compared with those of the shorter 18~cm core-pumped fiber in experiment H2014 \cite{hendersonsapir2014midinfrared}. This is due to the longer interaction length of this 2.8~m fiber and the lower power density of $\mathrm{P_1}$ in the core which prevents bleaching of the $\mathrm{^{4}I_{11/2}}$ (3) level.

\begin{figure}[!h]
	\begin{centering}
	\includegraphics[width=0.95\columnwidth]{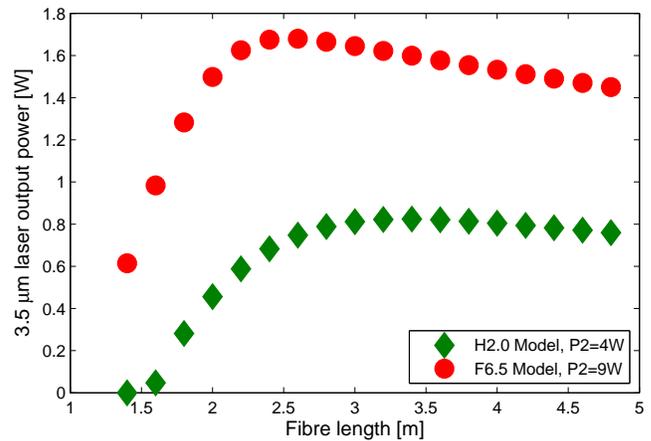}
	\par\end{centering}
	\caption{Modeled laser output power as function of fiber length based on parameters from experiments H2016 by Henderson-Sapir \textit{et al.}  \protect\cite{hendersonsapir2016versatile} (H2.0) and F2016 by Fortin \textit{et al.} \protect\cite{fortin2016watt} (F6.5). The H2.0 simulation had $\mathrm{P_2}$ power fixed at 4 W and the F6.5 simulation had $\mathrm{P_2}$ power fixed at 9 W.}
	\label{fig:Fiber length - Ori2016Laval2016}
\end{figure}

The modeled laser output power as a function of fiber length for fixed powers of both $\mathrm{P_1}$ and $\mathrm{P_2}$ is presented in Fig. \ref{fig:Fiber length - Ori2016Laval2016}. The model predicts an optimal fiber length of 3.4~m for the H2016 \cite{hendersonsapir2016versatile} H2.0 system when the second pump operates at 4~W. Laser power decreases sharply below 2.6~m interaction lengths and decreases moderately above 3.6~m. An optimal fiber length of 2.5~m is predicted for the F2016 \cite{fortin2016watt} F6.5 system when the second pump operates at 9~W. This implies a potential increase in laser power of 10\% compared with the experimental fiber length of 4.3~m. 

\begin{figure}[!h]
	\begin{centering}
	\includegraphics[width=1\columnwidth]{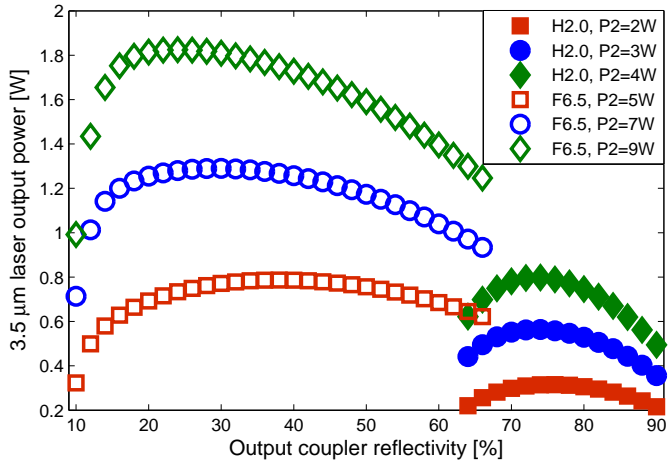}
	\par\end{centering}
	\caption{Modeled laser output power as a function of output coupler reflectivity based on parameters from experiments H2016 by Henderson-Sapir \textit{et al.} \protect\cite{hendersonsapir2016versatile} (H2.0) and Fortin \textit{et al.} \protect\cite{fortin2016watt} (F6.5). The H2.0 simulation had the power of $\mathrm{P_2}$ fixed at 4, 3, and 2~W. The F6.5 simulation had the power of $\mathrm{P_2}$ fixed at 9, 7, and 5~W.}
	\label{fig:Output coupler reflectivity - Ori2016Laval2016}
\end{figure}

Modeled laser output power as a function of output coupler reflectivity for fixed powers of $\mathrm{P_2}$ is presented in Fig. \ref{fig:Output coupler reflectivity - Ori2016Laval2016}. The plot shows how optimum output coupler reflectivity decreases with increasing power of the second pump. An optimal reflectivity of 74\% is predicted for the H2.0 \cite{hendersonsapir2016versatile} system when the second pump operates at 4~W and 76\% reflectivity at 2~W. An optimal reflectivity of 24\% is predicted for the F6.5 \cite{fortin2016watt} system when the second pump operates at 9~W. This implies a potential increase in laser power of 22\% compared with the experimental output coupler reflectivity of 55\%.

\section{Conclusions}

An extensive numerical model of DWP 3.5~\textmu m Er\textsuperscript{3+}-doped fiber lasers has been presented and validated against results from three published experiments. The model provides valuable insight into atomic and photonic interactions in both time and position along the fiber and enables the optimization of parameters such as fiber length, output coupler reflectivity, doping concentration, and pump wavelengths. The model may be extended to other dopant ions and fibers.

The limitations on DWP laser performance include the accumulation of ions in the lower laser level and the escape of ions from the second pump cycle. The dominant escape processes are the decay from the upper laser level to ground state and, in high Er\textsuperscript{3+}-doping concentrations, the cross relaxation process $W_{5362}$.

Future work includes FLAPP upgrade to account for laser wavelength shifting with second pump power and investigation into interionic processes by further modeling. Better understanding of interionic processes would enable us to improve optimization of doping and potential co-doping concentrations. We also intend to optimize the wavelength of the first pump for laser power and slope efficiency to achieve the optimum balance of ground and excited state absorptions.

\section*{Acknowledgments}

The authors would like to thank eRSA (South Australian provider of high performance computing) for the provision of computer resources.

\ifCLASSOPTIONcaptionsoff
  \newpage
\fi



%
%
%

\bibliographystyle{IEEEtran}
\bibliography{NumModDWP_arXiv}

%

\begin{IEEEbiography}[{\includegraphics[width=1in,height=1.25in,clip,keepaspectratio]{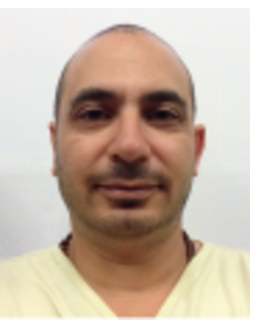}}]{Andrew Malouf} received his B.Sc. in experimental and theoretical physics from The University of Adelaide in 2013 and was awarded first class Honours in 2015. He is currently a Ph.D. candidate at The University of Adelaide with the Lasers and Optics Group.
\end{IEEEbiography}
\begin{IEEEbiography}[{\includegraphics[width=1in,height=1.25in,clip,keepaspectratio]{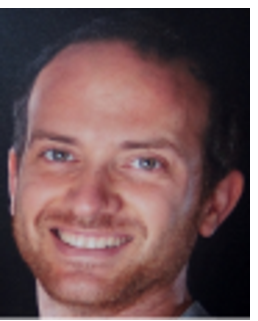}}]{Ori Henderson-Sapir} received his B.Sc. in physics and Mathematics from the Hebrew university in Jerusalem, Israel and his M.Eng. from the Tel-Aviv University in Israel where he worked on development of electrically small antennas. He recently received his Ph.D. from the University of Adelaide in Australia with the Lasers and Optics Group. His current research interests are mid-infrared sensing and the development of mid-infrared fiber lasers.
\end{IEEEbiography}
\begin{IEEEbiography}[{\includegraphics[width=1in,height=1.25in,clip,keepaspectratio]{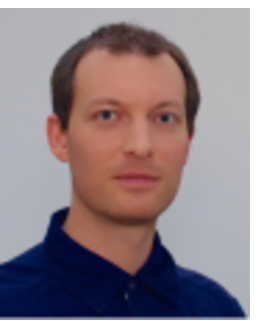}}]{Martin Gorjan} received his Ph.D. degree in physics in 2011 from University of Ljubljana for his work on the mid-infrared fiber lasers in collaboration with Fotona d.d. company in Ljubljana, Slovenia. In 2011 he joined McGill University, Montreal, Canada as a postdoc where he was researching mid-infrared lasers and non-linear phenomena. In 2012 he joined Max-Planck Institute for Quantum Optics, Munich, Germany as a postdoc to develop ultrafast thin-disk amplifiers and optical parametric conversion systems. In 2015 he joined R\&D department of Spectra-Physics, Rankweil, Austria where he continues to work on ultrafast lasers and amplifiers.  
\end{IEEEbiography}
\begin{IEEEbiography}[{\includegraphics[width=1in,height=1.25in,clip,keepaspectratio]{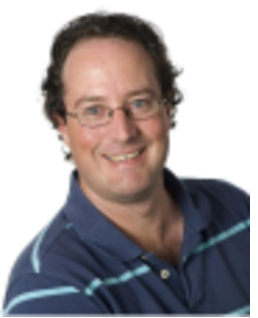}}]{David J. Ottaway} received the Ph.D. degree from The University of Adelaide, in 1999. His Ph.D. dissertation was on solid state laser sources for gravitational wave detection. In 2000, he joined the LIGO Laboratory, first as a Post-Doctoral Scholar at the LIGO Hanford Observatory, Richland, Washington, and later as a Staff Scientist with the Massachusetts Institute of Technology. During this period, he conducted research on commissioning the Initial LIGO detectors and developing optical and mechanical systems for the advanced LIGO detectors. In 2007, he joined as a Staff Member with The University of Adelaide, where he continues to develop laser and optical systems for advanced gravitational wave detectors and other forms of extreme remote sensing, including remote trace gas detection and atmospheric temperature studies.
\end{IEEEbiography}







\end{document}